\documentclass[onecolumn]{aa}
\usepackage{graphicx}
\usepackage{amsmath}
\usepackage{txfonts}
\usepackage{booktabs}
\usepackage{natbib}
\usepackage{nicefrac}
\bibpunct{(}{)}{;}{a}{}{,} % to follow the A&A style
\newcommand{\dg}{$^\circ$ }
\newcommand{\eq}[1]{\begin{equation}\label{eq:#1}}
\newcommand{\eeq}{\end{equation}}
\newcommand{\ccc}{CoRoT-7}

\renewcommand{\phi}{\varphi}

\numberwithin{equation}{section}

\begin{document}

\title{On the mutually inclined orbits of planets in the CoRoT-7 extrasolar
  planetary system}
 
\subtitle{CoRoT-7bcd}

\author{R. Dvorak\inst{1}, J. Schneider\inst{2}, V. Eybl\inst{1} }

\authorrunning{ Dvorak et al.}
\titlerunning{Inclined orbits in the \ccc\, system}

\offprints{R.\ Dvorak, \email dvorak@astro.univie.ac.at}

\institute{
Universit\"atssternwarte Wien, T\"urkenschanzstr. 17, A-1180 Wien, Austria
\and
LUTH, Observatoire de Paris-Meudon, 5 place J.Jansen, F-92195 Meudon, Paris, France
}

\date{Received -; accepted -}

\abstract  
{}
{We propose a method to be able to decide whether the
planets of CoRoT-7 are moving on mutually inclined orbits in the order
of  $i>10^{\circ}$.} 
{The extrasolar system CoRoT-7 is very special with
respect to the closeness of the planets to the host star, which results 
in a fast dynamical 
development. It would therefore be possible to determine the inclination of
the innermost planet CoRoT-7b with respect to the observer after 
an observation of at least three
years from space with the satellite CoRoT with sufficient precision.
Different inclinations would cause different duration of the transit times of
a planet in front of the star and would therefore give us a better knowledge
of the architecture of this system. With the aid of numerical integrations
and analytical estimations we checked how inclined orbits of additional
planets would change the transit duration of CoRoT-7b}
{After 3 years of observations when an additional planet would be on a
inclined orbit with respect to CoRoT-7b ($I_{mutual} > 10^{\circ}$) 
an increase of the order of minutes could be observed for the transit duration.   
}
{}

\maketitle{}

\section{Introduction}

Most of the more than 440 extrasolar planetary systems 
(=EPS)\footnote{see http://exoplanet.eu} are single planetary 
systems,  but as of October 2010 there are some 50 multiplanetary systems, where between 2 and 7 planets are known to orbit their host stars. A very special case is the
newly discovered system HD 10180 with seven planets orbiting an early G-type
star \citep{Lov10}. Certainly the small number of multiple EPS is a 
biased sample because it is highly improbable that there is just one planet around a star. From the theory of formation one expects that several planets may originate from the disk of gas and 
dust around a young star, like it is the case for the solar system.

An interesting fact is that many of the EPS-planets seem 
to move on large eccentric orbits and consequently 
-- when we expect more planets orbiting the star -- 
strong perturbations will act on the planets and therefore 
the  orbital elements may change significantly. Unfortunately 
our present methods to detect planets are just snapshots 
of the dynamical life time such that we don't have access to the dynamical 
evolution of a system. This is especially true for the detections via RV measuremenent 
where it is impossible to determine the inclinations of the orbital planes 
with respect to the observer. Even via transit observations of one planet 
combined with the measurement of radial velocities it is 
impossible to determine mutual inclinations 
of the orbits of planets when only one -- clearly the innermost -- is transiting.
The main problem to be solved would be to observe such an EPS for sufficiently long times to be able to see more than one snapshot in the dynamical evolution of the system. 
We are now in the same situation in astronomy as we had been during 
the last centuries when scientists wanted to determine the proper motion of stars 
or the orbital elements of double stars. Within the EPS most 
of the planets, especially the ones which we observe via transits, 
are relatively close to the host star and they have relatively large masses. 
So we might hope to see a signature of additional  
planets influencing the orbit of the inner planet which may result in a change 
of the transits times and/or the duration of the transit. Up to now it was impossible to make such precise 
measurements using ground-based observation but with the possibility of using satellites the situation changed.
The CoRoT spacecraft\footnote{The CoRoT space mission has
    been developed and is operated by CNES with the contribution of Austria,
Belgium, Brazil. ESA, Germany and Spain} (launched 2006) and the NASA Kepler mission  (launched 2009) permit very accurate measurement of light curves 
of transiting planets in terms of photometric precision and
timing of transits. The recent extension of the CoRoT mission for
another three years is important with respect to the observation of Transit Time Variations (TTV) and also possible Transit
Duration Variations (TDV) on a time line of several years.

\section{EPS with planets on inclined orbits}\label{sec:inclinedEPS}

The discovery of transiting planets with short periods makes
searches for additional planets in an EPS via TTV and TDV feasible. There are a number of
recent publications concerning  the possible perturbations in such systems
where  outer planets with eccentric or even inclined orbits with respect to the
innermost planet are present. 

\citet{Mar10} was investigating three of such
systems, namely HAT-P-13, HAT-P-7 and WASP-17. All three have orbital periods 
$2.2 \mbox{ days }< P < 3.7 \mbox{ days }$ and for HAT-P-13 an additional planets 
far away ($P = 448 \mbox{ days}$) on a large eccentric orbit ($e = 0.666$) is confirmed. 
The eccentricity of the
inner planet, assuming coplanar orbits, may suffer from larger variations due to
mutual inclinations of two planets, where the outer one is dominating the
angular momentum. This effect of amplitude variations is increasing with
increased mutual inclination.

An important effect when two planets are moving on inclined orbits is the
Kozai resonance which has a stabilizing effect for orbits with $40^{\circ} <
i_{mut} < 60^{\circ}$ having large eccentricities. 
In an article by \citet{Nag08} the effect of coupling between
mutual scattering, the Kozai mechanism and by tidal circularization was studied
with the aid of orbital integrations. They developed a theory that a planet
with small semimajor axis may not be formed by a type II migration because of
planet-disc interaction, but through an excitation of the eccentricity because
of the Kozai mechanism to
values close to $e = 1$ such that the pericenter is close to the host
star. Then the tidal friction may circularize the planet's orbit and as a consequence   
the authors predict that close-in-planets may have large inclinations up to
retrograde orbits.

\citet{Lib09b} treated the action of the Kozai 
resonance for five multiplanetary systems which are
not in Mean Motion Resonances (=MMR), namely $\upsilon$ Andromedae, HD 169830, HD 12661, HD 74156
and HD 155358. They found out by numerical studies in which they varied the
unknown inclinations and also the nodal longitudes, that with the exception of
HD 155358 all of them could be in the Kozai resonance when the mutual inclination is
larger than $45^{\circ}$. In another investigation by \citet{Lib09a} the role of mutually inclined orbits with respect to the stability of the orbits have been made in cases 
when the planets are in MMR. The authors have undertaken a parametric
study varying masses and orbital parameters of the planets where they also
took care of the migration rate and the rate of eccentriciy damping. It turned
out that due to a capture process in the early phases of the system formation
besides the 2/1 MMR (which was treated by \citealt{ThomLis}) also in the
3/1, 4/1 and 5/1 MMR the mutual inclinations may reach values as high as $70^{\circ}$ when 
the eccentricity of one planet is larger than $e > 0.4$. Due to their results they
say that \textsl{our simulations show that inclination excitation is a common
outcome, as long as eccentricity damping is not too strong}.  

The case of TrEs-2b was explored by \citet{Scu10}; this is an
EPS with a planet with a period $P = 2.4 \mbox{ days}$. Using older transit
observations from 2006 by \citet{Odo06} and comparing them
to the ones of the Kepler mission recently published by \citet{Gil10}
they could not find significant changes in the orbital parameters since the
discovery of that planet.  
  
In another system, GJ436, the transiting planet has a period of $P = 2.6 \mbox{ days }$
but a surprisingly large eccentricity of $e = 0.15$. \citet{Bat09} --
using secular evolution of a two and also a three planet system describing
GJ436 -- succeeded
in understanding this fact: the circularization has very long times scales when
other planets on eccentric orbits are present which may even move in almost
the same orbital plane as GJ436b.

A different topic in astronomy, namely the observation of period variations in
eclipsing binaries due to the presence of a third component, offers
possibilities of studying the so-called transit time variations (TTV) also in 
the analysis of transit photometry of extrasolar-planets \citep{Bor10}. But also the duration of an eclipse was studied in the
context of EB, and which is the consequence of an influence on the inclination
of the two components with respect to the line of sights. 
  
The subject of this article, the extrasolar planetary system CoRoT-7 is hosting one planet with the very small period of $P = 20 \mbox{ hours }$ and at least one other planet. We did numerical investigations for different sets of parameters which may cause a visible change
in the transit observations especially with respect to the possible change of
the inclination of the innermost planet CoRoT-7b.

\section{The extrasolar planetary system CoRoT-7}\label{sec:corot7system}

The EPS CoRoT-7 contains one transiting planet CoRoT-7b \citep{Leg09}
and one non transiting but confirmed planet CoRoT-7c detected by radial
velocity \citep{Que09}. 
A reanalysis of RV data led to the assumption that there could be a third planet in this
system \citep{Hat10}.  CoRoT-7 is -- up to now -- a unique example for an EPS where the dynamical 
time scales are very short which
is due to the closeness of the planets to the host star.
The short periods of 0.85, 3.7 and 9 days for the three planets mean that
the system is under very fast development: an integration time 
of a million years of CoRoT-7 in a numerical experiment is comparable to an 
integration of 100 million years for the solar system. Effects
caused by mutual perturbations of the planets will be visible on short time
scales. We list in Table \ref{tb:corotelements} the orbital elements of the CoRoT-7 planets, showing the results obtained from two different analyses of the data.

\begin{table}[h]
\caption{Orbital elements of the CoRoT-7 planets}
\begin{center}
\begin{tabular}{ l l l l}
%\hline 
\hline
  \noalign{\smallskip}
   & CoRoT-7b &  CoRoT-7c & CoRoT-7d \\
   Parameters &  \multicolumn{2}{c}{Data from \textsuperscript{1}} &\\
   \hline
\noalign{\smallskip}
\footnotesize
  mass [M$_{Earth}$] & 4.8$\pm$0.79 & 8.39$\pm$0.89 & \\
  a [AU]& 0.0172$\pm$0.00029 & 0.046 & \\
  P [days]& 0.853585$\pm$2.4$\times$10$^{-5}$ & 3.698$\pm$0.003 & \\
  e & 0 & 0 & \\
  i [degree]& 80.1$\pm$0.3 & - & \\
%  \hline
  \noalign{\smallskip}
  & \multicolumn{2}{c}{Data from \textsuperscript{2}} &\\
  \hline
 \noalign{\smallskip}
  mass [M$_{Earth}$] & 6.9$\pm$1.4 & 12.4$\pm$0.42 & 16.7$\pm$0.42 \\
  a [AU]& 0.017 & 0.045 & 0.08 \\
  P [days]& 0.853589 & 3.691 & 9.021\\
  & $\pm$0.00059 & $\pm$0.0036 & $\pm$0.019\\
  e & 0.0 & 0.080$\pm$0.050 & 0.0$\pm$0.05 \\
  i [degree]& 80$\pm$0.3 \footnotesize{\textsuperscript{3}}& $\le$85 \footnotesize{\textsuperscript{3}} & - \\
  \hline
  \noalign{\smallskip}\noalign{\smallskip}
\multicolumn{4}{l}{\footnotesize{\textsuperscript{1} http://exoplanet.eu}} \\
\multicolumn{4}{l}{\footnotesize{\textsuperscript{2} \citet{Hat10}}}\\
\multicolumn{4}{l}{\footnotesize{\textsuperscript{3} \citet{Que09}}}\\
\end{tabular}
\end{center}
\label{tb:corotelements}
\end{table}

\begin{figure}
\label{fig:31abcd}
\centering
\includegraphics[width=3.46 in,angle=270]{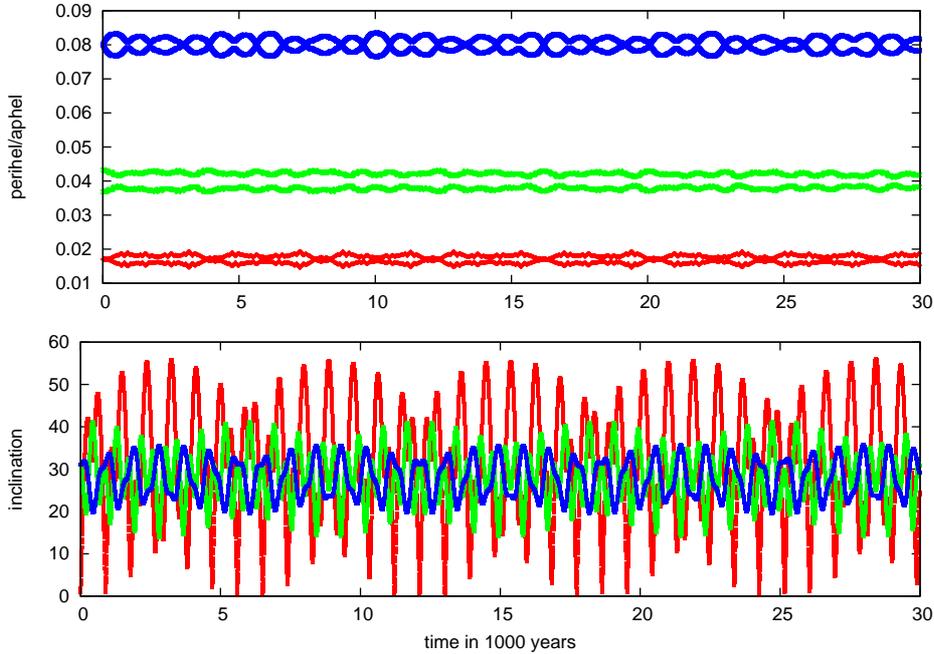}
\caption{Change in the perihelion and aphelion distances (upper graph) and inclinations (lower graph) for the \ccc\, planets with initial inclinations $i_{7c}= i_{7d}= 30^{\circ}$ and $i_{7b}=1^{\circ}$ during 30,000 years  (for colour figures see the online version of this journal).}
%\end{center}
\end{figure}

\section{Investigation of mutually inclined planets in the CoRoT-7 system}\label{numericalexp}

Former estimations showed that the regions where the \ccc\, planets are moving
are very stable \citep[e.g.][]{Hat10}. We show that even inclinations of
$i_{7cb} \sim 60^{\circ}$ have no big influence on the eccentricity of the
planets which would make the system unstable (see Fig. \ref{fig:31abcd}). The eccentricity changes are well inside the probable errors in the derived
element eccentricity. The situation is completely different however for the inclination,
because the duration of a transit is very  sensitive to it. According to the
published values of the inclination, respectively the errors 
($i = 80.1 \pm 0.3 $\dg), a shift of several tenth of degrees should be observable. In this section we will describe the numerical experiments and the results we obtained. 

\begin{figure}
\centering
\includegraphics[width=3.46in,angle=270]{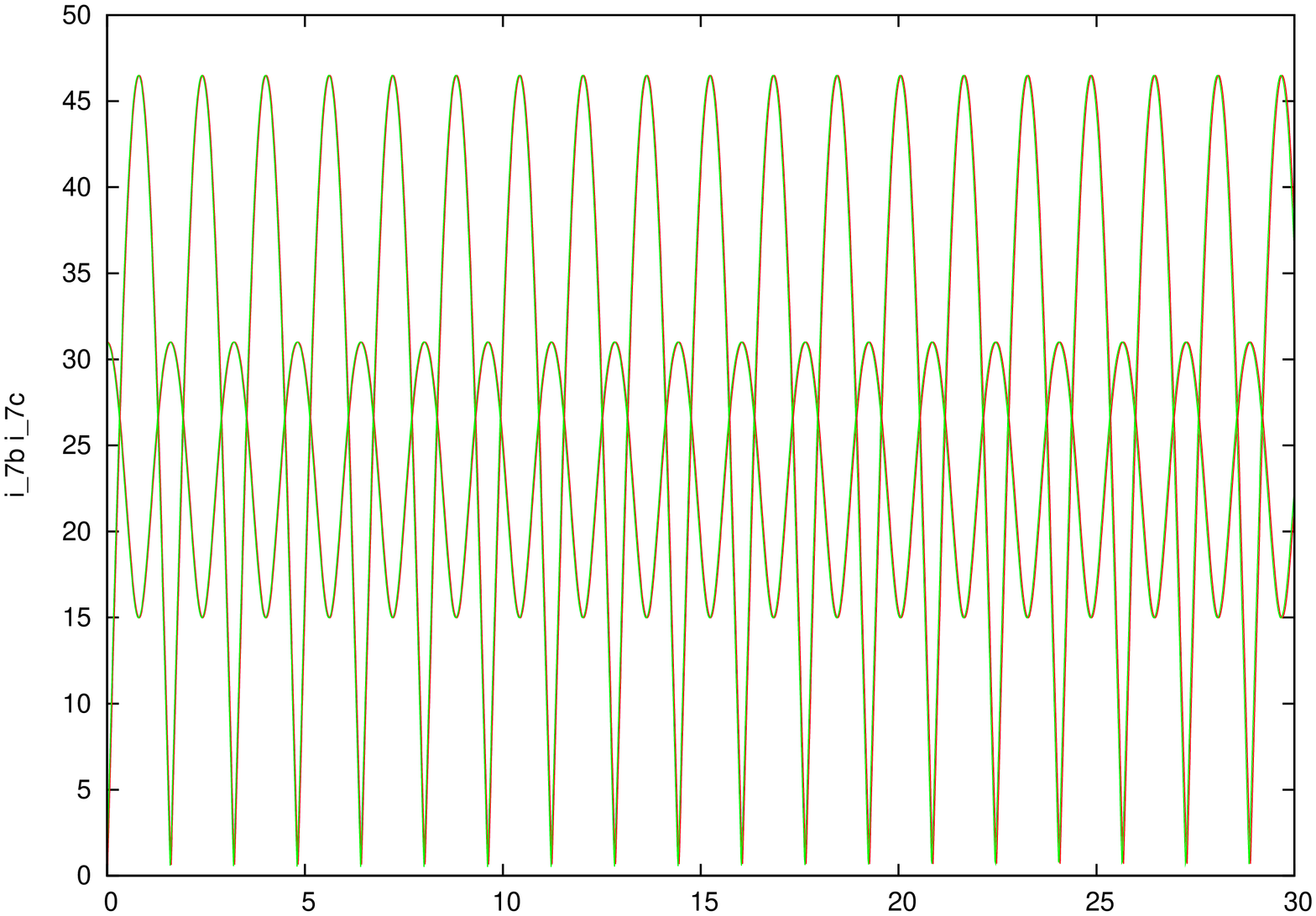}
\caption{Evolution of the inclination for  CoRoT-7b and CoRoT-7c during 30,000
  years of  integration with initial inclinations $i_{7b}=1^{\circ}$ and
$i_{7c}=30^{\circ}$. There is no visible difference between {\sf Runs A} and {\sf B} (for details see text).} 
\label{fig:31a}
%\end{center}
\end{figure}

\begin{figure}
\centering
\includegraphics[width=3.46in,angle=270]{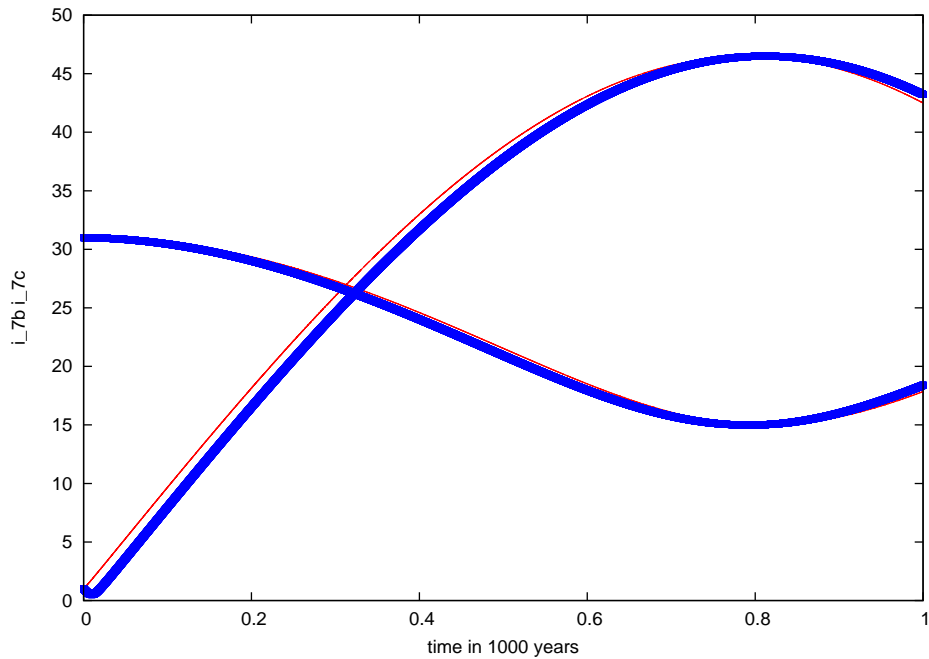}
\caption{Evolution of the inclination for  CoRoT-7b and  CoRoT-7c during the first 1,000
  years of  integration with initial inclinations $i_{7b}=1^{\circ}$ and
$i_{7c}=30^{\circ}$. {\sf Run A} and {\sf Run B} are shown as thin and thick lines, respectively.}
\label{fig:31b}
%\end{center}
\end{figure}

\begin{figure}
\centering
\includegraphics[width=3.46in,angle=270]{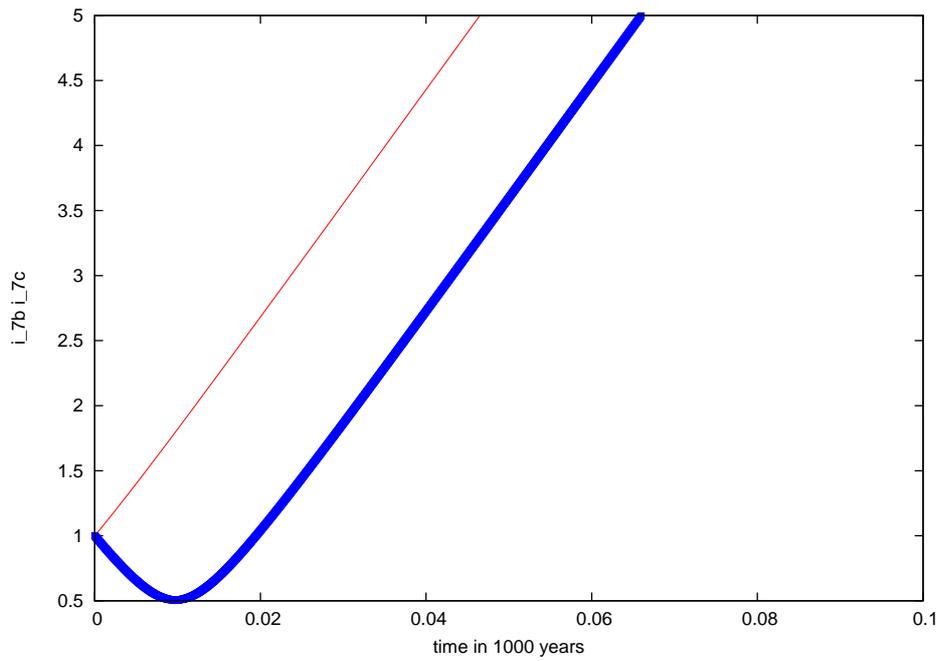}
\caption{Detail of Fig. \ref{fig:31b}: Evolution of the inclination for  CoRoT-7b and  CoRoT-7c during the first 100
  years of  integration. {\sf Run A} and {\sf Run B} are shown as thin and thick lines, respectively.}
\label{fig:31c}
%\end{center}
\end{figure}

\begin{figure}
\centering
\includegraphics[width=3.46in,angle=270]{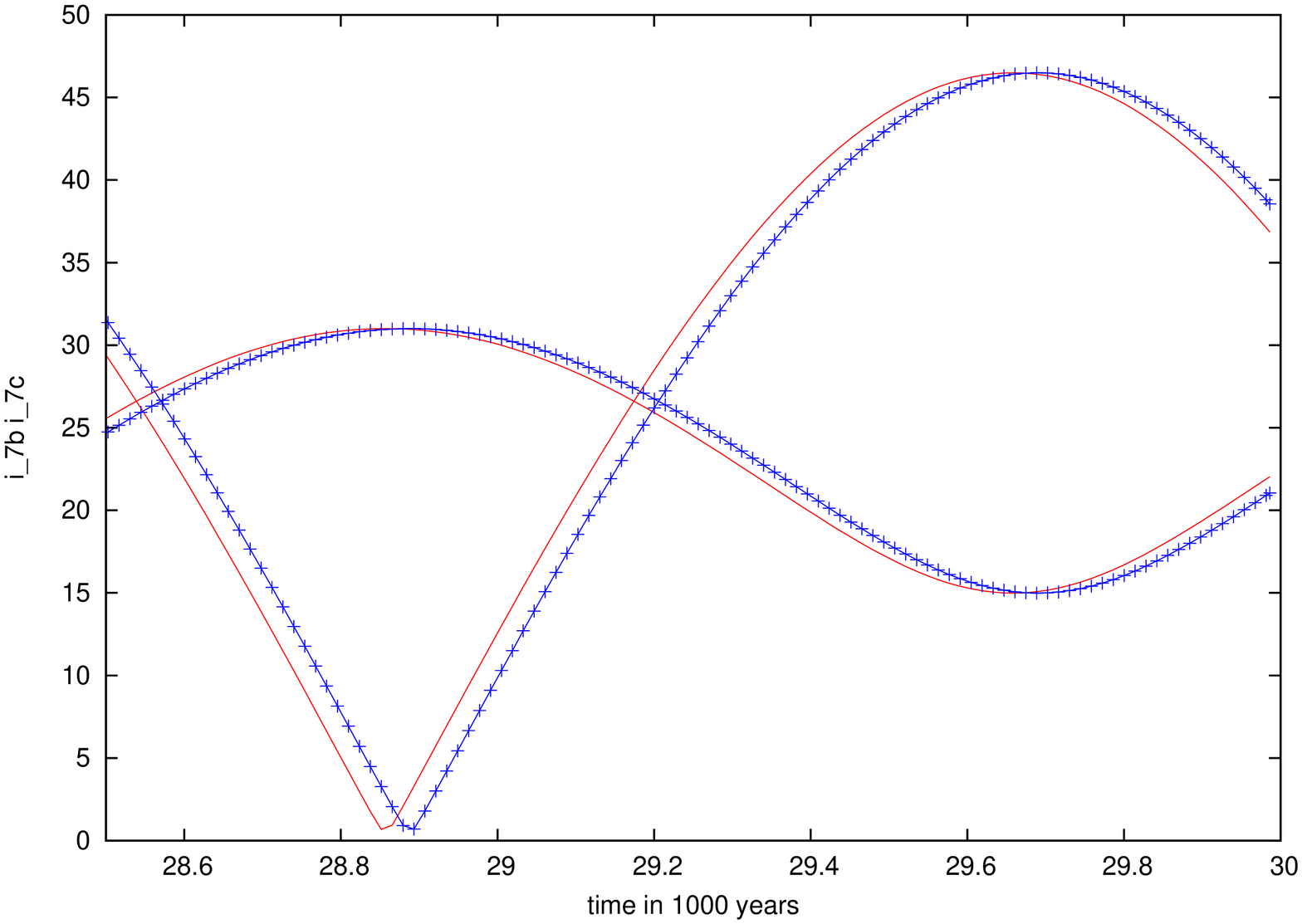}
\caption{Evolution of the inclination for CoRoT-7b and CoRoT-7c for the last 400 years of the 30,000 year integration time. {\sf Run A} and {\sf Run B} are shown as solid and crossed lines, respectively.}
\label{fig:31d}
%\end{center}
\end{figure}

It is particularly interesting whether one could
determine how much the transiting planet would be influenced with respect to
its inclination; this would have consequences for the transit time and the
transit time duration. Therefore
different test computations have been undertaken numerically (see Section \ref{sec:numerical2}) as well as using an analytical approach (see Section \ref{sec:analytical}) for inclined planets in the \ccc\, system. 

\subsection{Numerical results for the two planet case}\label{sec:numerical2}
In this section we considered only the two confirmed planets in the \ccc\, system; an extension to the possible case of three planets can be found in Section \ref{sec:numerical3}.
In Figs. \ref{fig:31a}, \ref{fig:31b}, \ref{fig:31c} and \ref{fig:31d} we show the effect on
the inclination of CoRoT-7b caused by a difference in the  ascending node ($\Omega_{7b}=20^{\circ}$ and 
$\Omega_{7c}=140^{\circ}$ = {\sf Run A}, respectively vice
versa = {\sf Run B}). For each run, the following initial inclinations were used: $i_{7b}=1^{\circ}$ and
$i_{7c}=30^{\circ}$, the time scale for the integration was $3.10^4$ years.

In Fig. \ref{fig:31a} there is no visible difference between {\sf Run A} and {\sf
  Run B} over a time period of 30,000 years; note that \ccc b reaches inclinations up to
$i_{7b}=47^{\circ}$. A closer look at a shorter period of time
(1,000 years) shows a slight time shift for the evolution of the inclination of
\ccc b (Fig. \ref{fig:31b}). Zooming into the first 100 years of the integration it turns out that the time shift is caused by  a different behaviour during the first ten years:
while $i_{7b}$ increases immediately in {\sf Run A}, in {\sf Run B} it decreases at first, but starts to increase after about ten years. From this point on the dynamical behaviour is qualitatively and quantitatively the same.: The thick and thin line in Fig. \ref{fig:31c} show the same slope after the initial ten years. This time shift remains during the whole time which can be seen in Fig. \ref{fig:31d},  where we plotted the respective evolution of the inclinations after several thousand years.

\subsection{Analytical estimation}\label{sec:analytical}
  
We analytically estimate the change of the orbital elements in the two planet case (subscripts {\sf i} and {\sf
  j})using the differential equations for the secular perturbation of the orbital elements $e,\omega,i,\Omega$ from \citet{Stu65}:
\begin{subequations}\label{eq:secdiff}
\begin{align}
\label{eq:sec1}
\frac{\mathrm{d} e_{i}}{\mathrm{d} t}\, =&\, \frac{ d_{i}}{ e_{i}} \, \left ( f_{i}\, g_{j}-g_{i}f_{j} \right )\, =\, d_{i}\, e_{j}\, \sin \left ( \omega_{j}-\omega_{i} \right ) \\
\label{eq:sec2}
\cos i_{i} \, \frac{\mathrm{d} i_{i}}{\mathrm{d} t}\, =&\, c_{i}\, \sin i_{j} \, \sin \left ( \Omega_{i}-\Omega_{j} \right ) \\
\label{eq:sec3}
 \frac{\mathrm{d} \omega_{i}}{\mathrm{d} t}\,=&\, c_{i}-d_{i}\, \frac{e_{j}}{e_{i}}\, \cos\left ( \omega_{j}-\omega{i} \right )\\
\label{eq:sec4}
 \frac{\mathrm{d} \Omega_{i}}{\mathrm{d} t}\,=&\, -c_{i} \, \left \{ 1-\frac{\sin i_{j}}{\sin i_{i}} \, \cos \left (  \Omega_{i}-\Omega_{j}\right )\right \}
\end{align}
\end{subequations}

In the case of a heavy central body (the star) and two planets, the right side
of equations \ref{eq:secdiff}
can be assumed to be constant ignoring the mutual periodic perturbations. 
The orbital elements $e,\omega,i,\Omega$ can then be described as 
slowly increasing or decreasing functions of time. The differential equation 
for the inclination \ref{eq:sec2} can thus be written as

\begin{equation}\label{eq:deltai}
\Delta i_{i} = - c_{i} \,  \sin \left ( \Omega_{j}-\Omega_{i} \right ) \, \frac{\sin i_{j}}{\cos i_{i}}
\end{equation}

depending only on the current values of $i_{i},i_{j},\Omega_{i}$ and
$\Omega_{j}$. 
The $c_{i}$ are described by

\begin{equation}\label{eq:ci}
c_{i}=\frac{1}{4}\, m_{j} \, n_{i} \, \, b_{1}^{(\nicefrac{3}{2})}
\end{equation}

where $b_{1}^{(\nicefrac{3}{2})}$ is the Laplace coefficient of the first order. The $b_{1}^{(\nicefrac{3}{2})}$ depend on the ratio of the orbital distance of the two planets $\alpha$ and are calculated as follows:

\eq{b1}
b_{k}^{(s)}=\tfrac{2}{\pi } \int_{0}^{\pi}\frac{\cos k \gamma}{(1-2\alpha \cos\gamma +\alpha ^{2})^{s}}\: \mathrm{d}\gamma ,
\eeq

with $\alpha=\frac{a_{i}}{a_{j}}$.

\begin{figure}
\centering
\includegraphics[width=3.46in]{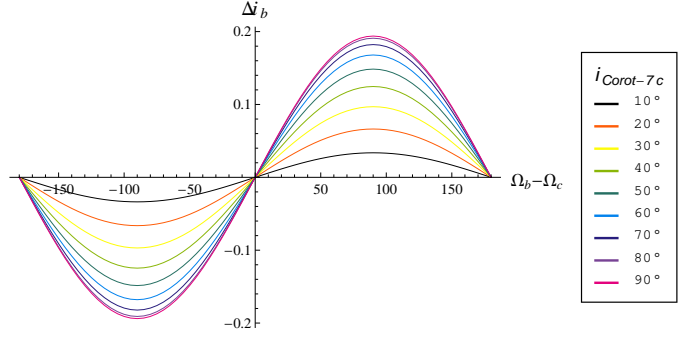}
\caption{Secular change in the inclination of \ccc b per year as a function of the relative position of the ascending node and the inclination of \ccc c}
\label{fig:deltai1}
%\end{center} 
\end{figure}

\begin{figure}
\centering
\includegraphics[width=3.46in]{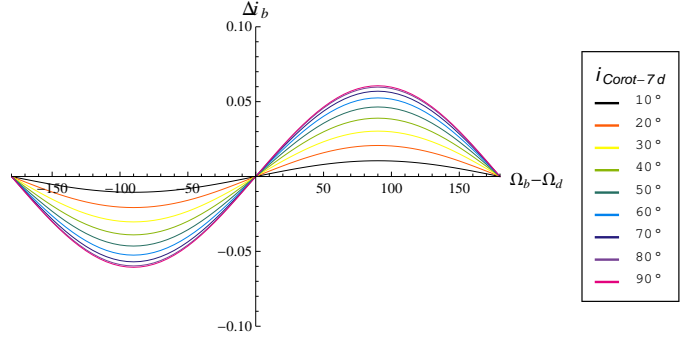}
\caption{Secular change in the inclination of \ccc b per year as a function of the relative position of the ascending node and the inclination of \ccc d (note the difference in scale to Fig. \ref{fig:deltai1})}
\label{fig:deltai2}
%\end{center}
 \end{figure}

Thus, for a given set of $i_{i},i_{j},\Omega_{i}$ and $\Omega_{j}$ it is 
possible to estimate the change $\Delta i$ using \ref{eq:deltai}. 
Figs. \ref{fig:deltai1} and \ref{fig:deltai2} show the secular change in
inclination of \ccc b in 
degrees per year as a function of the relative position of the ascending node 
$\Delta \Omega = \Omega_{j}-\Omega_{i}$ and the inclination of the other
planet. If we take into account only the perturbation of planet \ccc c on \ccc b,
the change in its inclination occurs at a rate of $\pm 0.2$\dg per year at a separation of
the ascending nodes of $\pm 90$\dg. Considering only the perturbation of \ccc d,
the change is much smaller, namely $\pm 0.06$\dg per year (see Figs. \ref{fig:deltai1} and \ref{fig:deltai2}). The initial value for the inclination of \ccc b was assumed to be 1\dg in both calculations.

\begin{figure}
\centering
\includegraphics[width=3.46in]{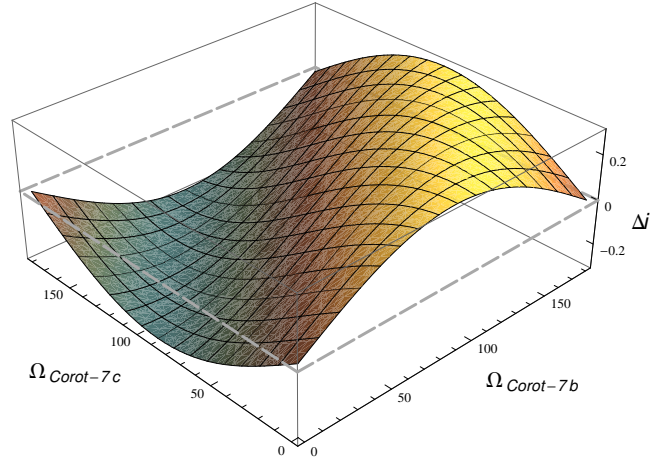}
\caption{Change in the inclination of \ccc b after 3 years as a function of the
  relative position of the ascending node of \ccc b and \ccc c using equation \ref{eq:deltai}.}
\label{fig:deltai3d}
%\end{center}
 \end{figure}

\begin{figure}
\centering
\includegraphics[width=3.46in,angle=270]{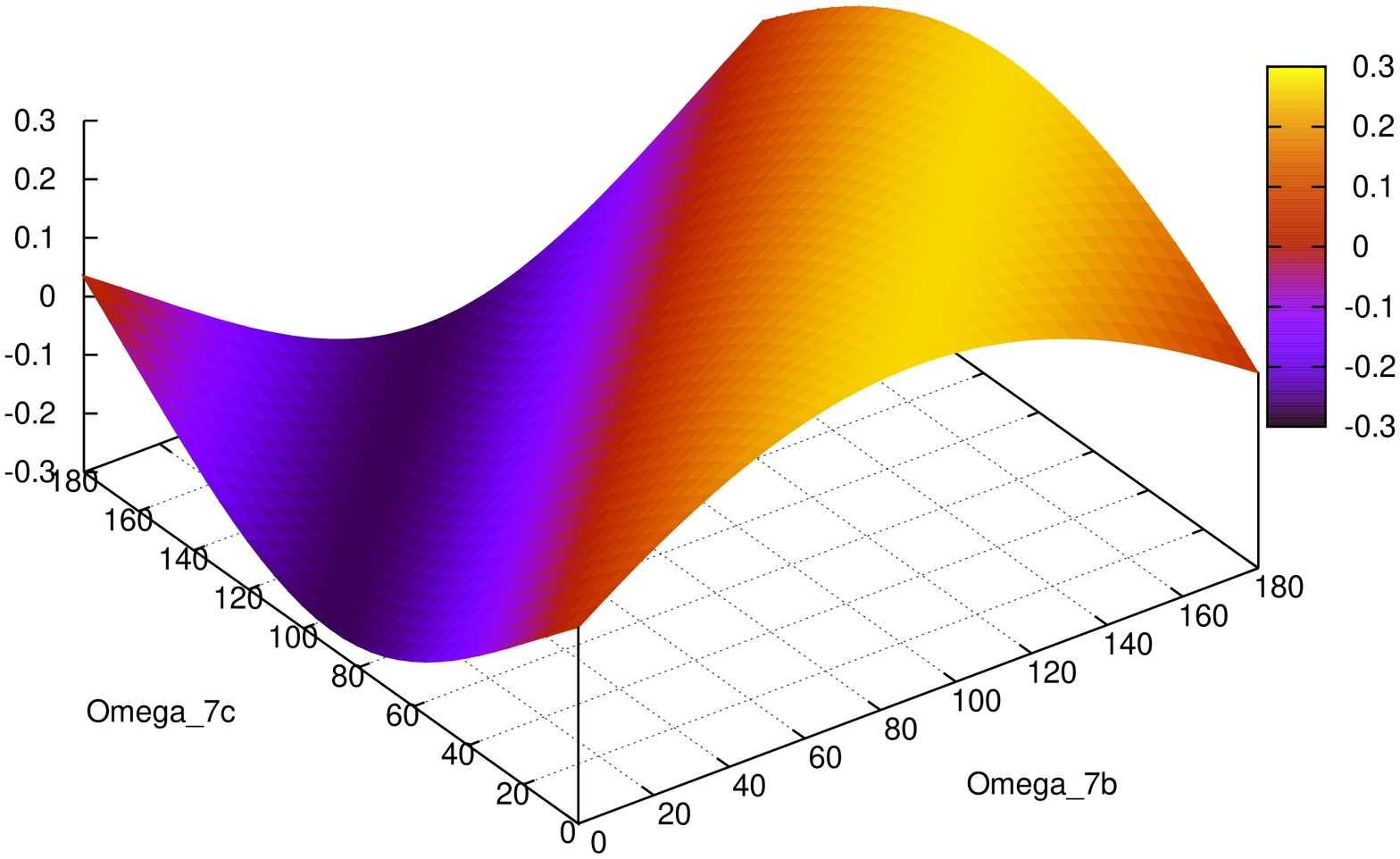}
\caption{Change in the inclination of \ccc b after 3 years as a function of the
  relative position of the ascending node of \ccc b and \ccc c using results
  from numerical integrations}
\label{fig:3y-31}
%\end{center} 
\end{figure}

The results from this analytical approach fit the data from numerical integration very well, as can be seen in the following graphs: Figs. \ref{fig:deltai3d} (analytical model) and \ref{fig:3y-31} (numerical model) show the change in $\Delta i_{7b}$ after 3 years as a function of $\Omega_{7c}$ and $\Omega_{7d}$, again with an initial inclination $i_{7b}$ = 1\dg.

\subsection{Extension to the three planet case}\label{sec:numerical3}
  
Since the data analysis of \citet{Hat10} lead to the conclusion that (at
least) three planets exist in the extrasolar planetary system \ccc, we investigated the combined
influence of different initial conditions for  \ccc c and \ccc d on the
inclination of the transiting planet. Here we ignored the possible differences in the ascending nodes
of the three planets and set all to the same value. This assumption may be
somewhat artificial but the parameter space is so large that this is the only
way to estimate possible changes of the $i_{7c}$. We varied the inclinations of both outer planets from normal prograde to retrograde
orbits ($0^{\circ}<i<180^{\circ}$). 

\begin{figure}
\centering
\includegraphics[width=3.46in,angle=270]{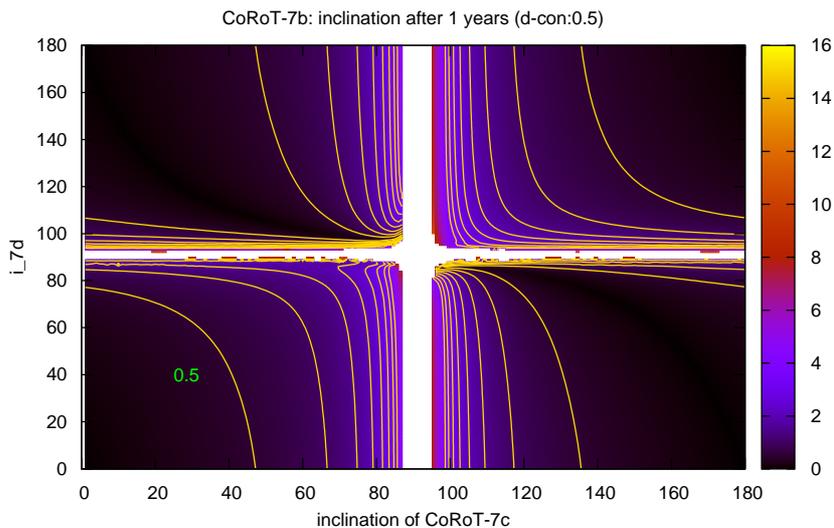}
\caption{Change in the inclination of \ccc b per year as a function of the
  relative position of the ascending node of \ccc b and \ccc c using results
  from numerical integrations.}
\label{fig:1y}
%\end{center}
 \end{figure}

\begin{figure}
\centering
\includegraphics[width=3.46in,angle=270]{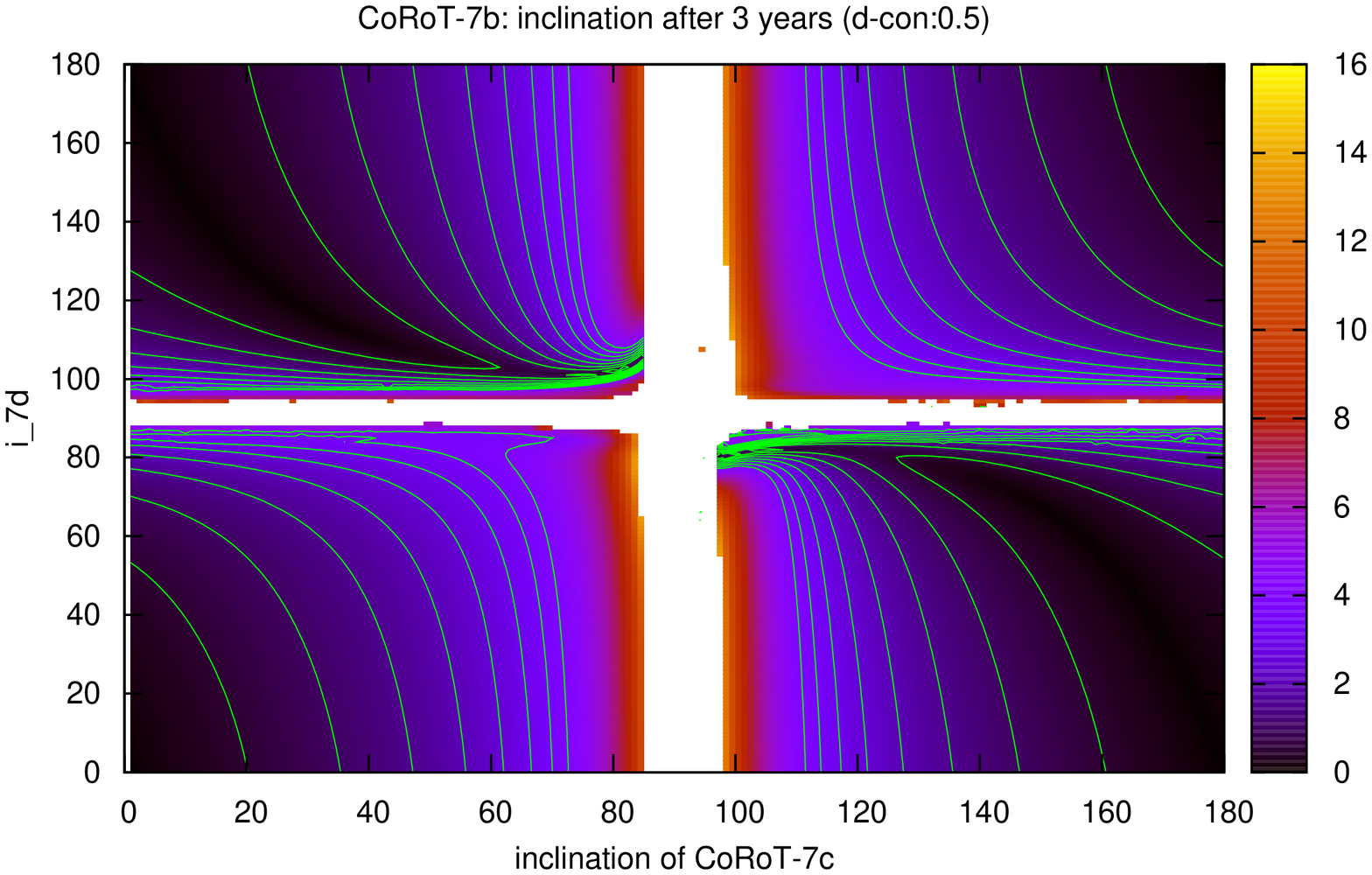}
\caption{Change in the inclination of \ccc b per year as a function of the
  relative position of the ascending node of \ccc b and \ccc c using results
  from numerical integrations.}
\label{fig:3y}
%\end{center}
 \end{figure}

\begin{figure}
\centering
\includegraphics[width=3.46in,angle=270]{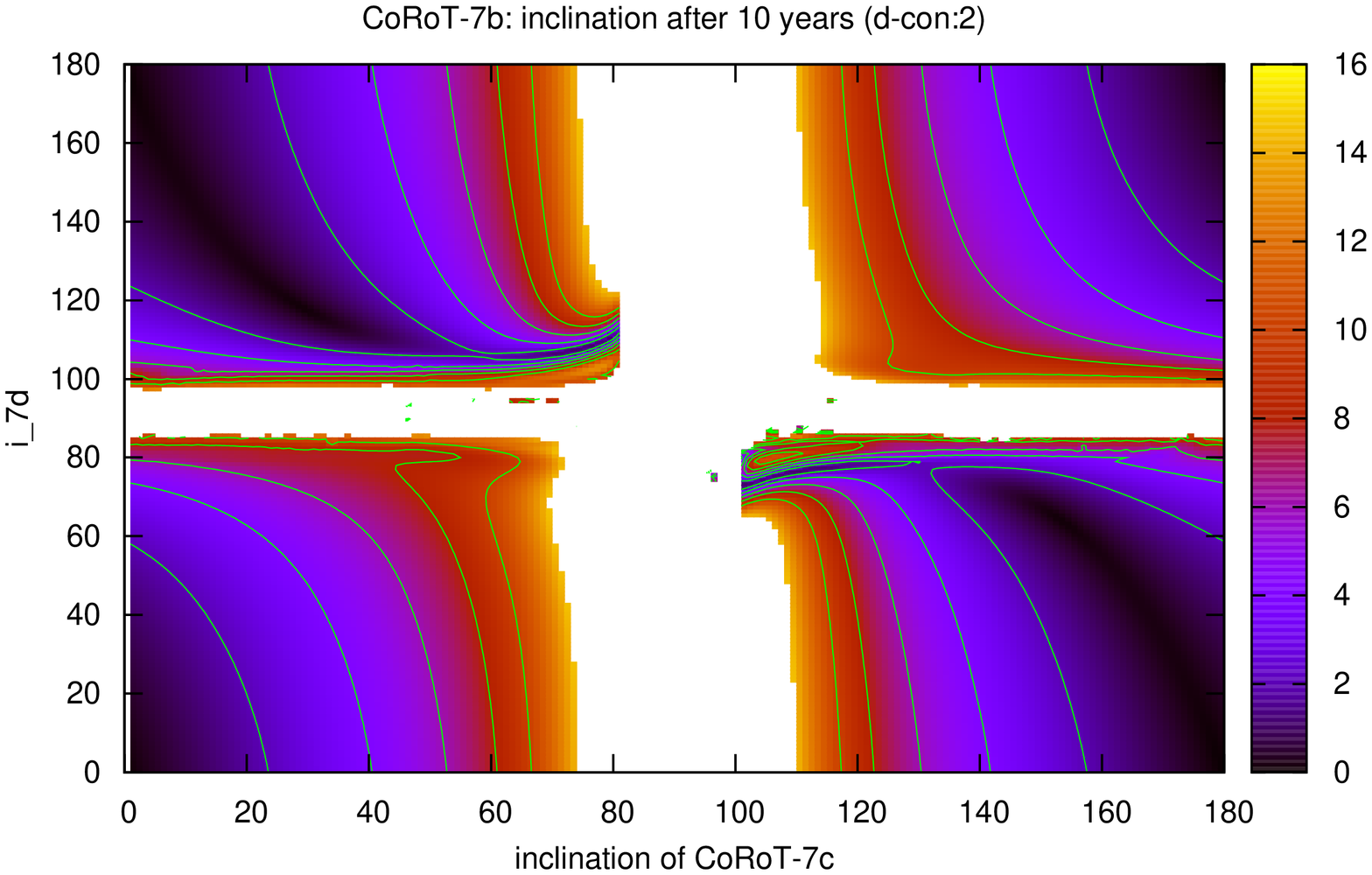}
\caption{Change in the inclination of \ccc b per year as a function of the
  relative position of the ascending node of \ccc b and \ccc c using results
  from numerical integrations.}
\label{fig:10y}
%\end{center} 
\end{figure}

\begin{itemize}

\item {\sf After 1 year}: The results of the integration for such a short time interval (Fig. \ref{fig:1y}) show that only for large inclinations of \ccc c ($i \sim
60^{\circ}$) \ccc b would suffer from changes in the order of $1^{\circ}$
without a big influence of the inclination of \ccc d. On the other hand the
outermost planet would have a significant influence only if its orbital plane was almost perpendicular to the plane of \ccc b. This is also clear from the
results of analytical computations (see Figs. \ref{fig:deltai1} and \ref{fig:deltai2}).

\item {\sf After 3 years:} During the lifetime of the CoRoT satellite the change in the
  inclination of \ccc b would be observable within the given error bars, when both outer planets would have inclinations in the order of 
 several tenths of degrees (Fig. \ref{fig:3y}). 

\item {\sf After 10 years:} It is evident from Fig. \ref{fig:10y} that for such a long time
  interval of observations the changes could be significant even for only slightly
  inclined orbits of one or both of the outer planets. 

\end{itemize}

The cross-shaped white area (for retrograde orbits) of the parameter space would
lead to unstable orbits even in short time scales.

\section{Transit duration}\label{sec:transitduration}

Several formulas have been published to calculate the duration of
transits \citep[e.g.][]{Kip08,Sea03}. To determine the transit duration of \ccc b for 
different apparent inclinations we applied the following formula of \citet*{Sea03}

\eq{smo}
t_{T}\,=\,\frac{P}{\pi}\, \arcsin \left( \frac{R_{*}}{a} \left[ \frac{ \left( 1+\frac{R_{P}}{R_{*}} \right) ^2\,-\, \left( \frac{a}{R_*}\,\cos i)^2 \right)}{1\,-\,\cos^2 i} \right]^{1/2} \right),
\eeq
being valid only for circular orbits. Note that in our calculations we always consider the total transit duration, i.e. including ingress and egress of the planetary disk. 

The numerical integrations of the \ccc\,system for ten years (see Fig. \ref{fig:10y}) resulted in a change in the inclination of \ccc b of up to 3\dg. Table \ref{tb:ttrans} shows the corresponding change in the transit duration.
%\textbf{Note, that the value for the transit duration given in Table \ref{tb:ttrans} from L{\'e}ger et al does not correspond to the value calculated with formula
%\ref{eq:smo}, which amounts to approx. 67 minutes! $\Rightarrow$ problem?! Should this be mentioned in the paper?}

\begin{table*}[ht]
\begin{center}
\footnotesize
\begin{tabular}{l l  l l l l l l l l l l l l l}
\hline
\noalign{\smallskip}
\multicolumn{1}{l}{\emph{i} \emph{[degree]}} & 77.1& 77.6& 78.1& \multicolumn{1}{c}{78.6}& 79.1& 79.6& 80.1 \footnotesize{\textsuperscript{1}} & 80.6& \multicolumn{1}{c}{81.1}& 81.6 &  82.1& 82.6& \multicolumn{1}{c}{83.1}\\
\noalign{\smallskip}
\multicolumn{1}{l}{\emph{$\Delta t_{trans}$} \emph{[min]}} & 34.5 & 42.3 & 48.6 & \multicolumn{1}{c}{53.9} & 58.6 & 62.7 & 66.4 &
69.7 & \multicolumn{1}{c}{72.7} & 75.4 & 77.8 & 80.1 & \multicolumn{1}{c}{82.1} \\
\hline
%\noalign{\smallskip}
%\multicolumn{1}{c }{}&\multicolumn{4}{c|}{}&\multicolumn{5}{c|}{\sc over 3 years}&\multicolumn{4}{c}{}\\
%\cline{6-10}
%\noalign{\smallskip}
%\multicolumn{1}{c|}{}&\multicolumn{13}{c|}{\sc over 10 years}\\
%\cline{2-14}
\noalign{\smallskip}\noalign{\smallskip}\noalign{\smallskip}
\multicolumn{14}{l}{\footnotesize{\textsuperscript{1} currently adopted value of inclination (http://exoplanet.eu)}}
%\multicolumn{14}{l}{\footnotesize{\textsuperscript{2}}
\end{tabular}
\end{center}
\caption{Change in the transit duration of \ccc b.}
\label{tb:ttrans}
\end{table*}

The total duration of the transit, as calculated with Formula \ref{eq:smo} -- which in this case exactly describes the transit geometry -- is 66.4 minutes. This is in very good agreement with the transit duration published by \citet{Leg09} of $1.125 \pm 0.05 $h $= 67.5 \pm 3$ min.
If the apparent inclination of \ccc b decreases by a value of 3\dg, the transit duration decreases by approx. 32 minutes, whereas if the inclination increases by 3\dg the transit duration is approx. 16 minutes longer (see Table \ref{tb:ttrans}). This could be easily observed by the CoRoT satellite. Our numerical calculations suggest -- depending on the exact conditions -- that a 1\dg change in the inclination of \ccc b could be expected after three years, and a 3\dg change should be possible over the course of 10 years (see Section \ref{sec:numerical3}).
One a side note, if the apparent inclination of \ccc b is less than 76.2\dg the transit cannot be observed any more.

Since CoRoT-7 is a multiplanetary system with at least two planets -- as already pointed
out (\citealt{Que09,Hat10};  see Table \ref{tb:corotelements}) -- we also investigated the possibility of \ccc c or \ccc d  becoming a transiting planet. Taking the orbital plane of \ccc b to be the
reference plane of the system, we calculated the transit possibility as a function of
inclination and longitude of the ascending node in reference to \ccc b. These calculations
could not be done using formula \ref{eq:smo}, which takes into account only
the apparent inclination, but not the longitude of the ascending node. This is
an important orbital parameter for determining the transit probability.
We adopted a semi-analytical approach using the computer algebra software 
\textit{Wolfram Mathematica$^{TM}$}. The planetary orbit is represented by a 3D
parametric equation. The full representation of the orbit is obtained by
combining a parametrization of the position of the orbital plane according to
the Keplerian elements $e,a,i,\omega$ and $\Omega$, with a rotation of the 
system according to the point of view of the observer. This approach allows a
consistent choice of the reference plane for multiplanetary systems, and
accounts for the apparent inclination the system is observed under. We use an extended
projection of the stellar disk in direction of the observer, represented by
a cylindrical parametrization with a diameter of $R=R_*+R_P$, so that we get the total transit duration. The intersection 
of the orbit with the projection of the disk yields two positions $t_1$ and
$t_2$ on the orbital curve, which correspond to the starting and ending points 
of the transit.
For a circular orbit, which we assume for the planets of the \ccc\, system, the parameter $t$ of the orbital curve is equivalent to 
the mean motion of the planet. In order to obtain the amount of time passing 
between the planet entering and exiting the stellar disk we calculate the 
length $l$ of the arc of the orbital curve between the two points $t_1$ and $t_2$

\eq{arc}
l\,=\,2\pi\,a\,\frac{t_2\,-\,t_1}{360^{\circ}}
\eeq

The ratio of the orbital length during the transit to the circumference of the
orbit $L$ corresponds to the ratio of the duration of the transit to the 
planetary period $P$. The transit duration can then be written as

\eq{ttrans}
t_{Transit}\,=\,\frac{l}{L}\,P.
\eeq

\begin{figure}
\centering
\includegraphics[width=3.46in]{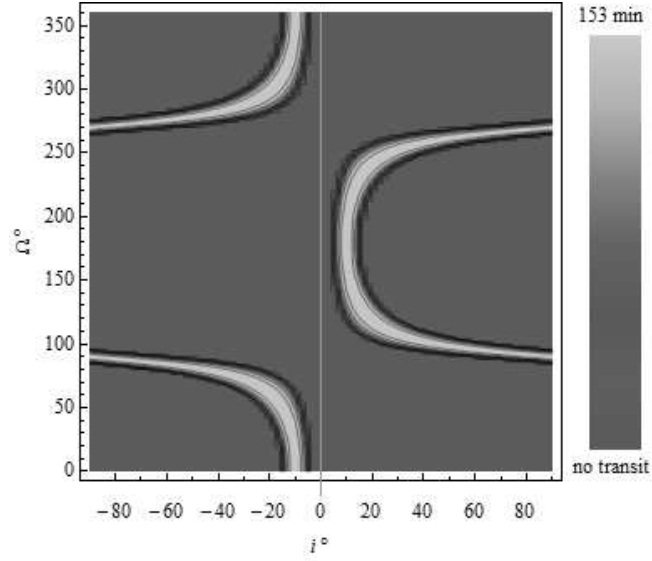}
\caption{Duration of a possible transit of \ccc c as a function of inclination
  and longitude of the node. The transit duration is given in minutes and indicated by color coding and contour lines.}
\label{fig:transitc}
%\end{center}
\end{figure}

\begin{figure}
\centering
\includegraphics[width=3.46in]{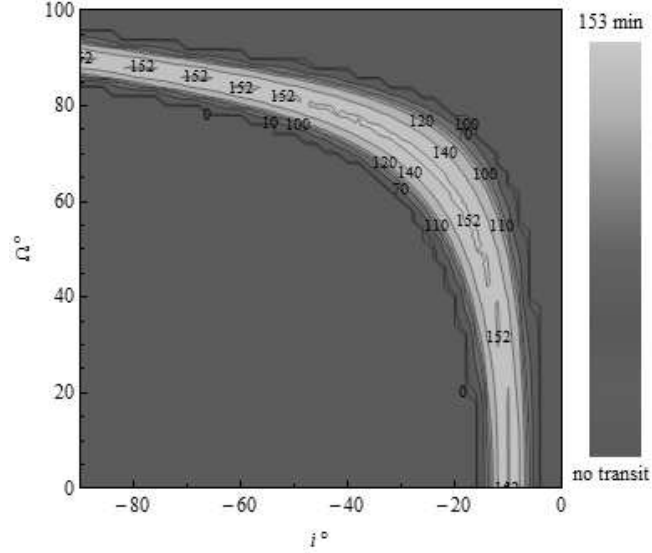}
\caption{Detail of Fig. \ref{fig:transitc} with labeled contour lines.}
\label{fig:parttransitc}
%\end{center}
 \end{figure}

For investigating the possibility of \ccc c or \ccc d becoming a transiting planet, we had to adopt a value for the planetary radii which are currently not known. 
\ccc c has a minimum mass of $8.4\,M_{Earth}$, positioning it right between Earth and Neptune. So we assumed the density of \ccc c to be the mean value of the densities of Earth and Neptune, yielding a radius of $0.22\,R_{Jupiter}$. \ccc d has a minimum mass very similar to the mass of Neptune ($16.5 M_{Earth}$), so it was assumed to have the same density, corresponding to a radius of $0.34\,R_{Jupiter}$.
 Figures \ref{fig:transitc} and \ref{fig:parttransitc} show the duration of a transit of \ccc c as a function
of its inclination and the position of the ascending node. The reference plane 
is fixed to the plane of \ccc b and the ascending node of \ccc b is assumed 
to be in the direction of the observer. Thus, the value of the inclination 
as shown in Fig. \ref{fig:transitc} is not the apparent inclination as it 
would appear to an observer, but the in-system inclination with respect to 
the plane of \ccc b. Fig. \ref{fig:parttransitc} is showing a 
enlarged detail of Fig. \ref{fig:transitc} with labeled contours of equal 
transit duration. The same calculation was done for \ccc d, and qualitatively 
the same behaviour was found. Due to the similarity of the plots -- the only difference being in the scale -- we refrain 
from showing this figure. The maximum transit duration is 153 minutes 
for \ccc c, as can be seen in Figs. \ref{fig:transitc} and \ref{fig:parttransitc}, and 217 minutes for \ccc d.

\section{Conclusions}
 
In this investigation we estimated the TDV   
caused by inclined additional planets which reaches -- although relatively small --
values of some degrees and is therefore detectable with the CoRoT satellite. 
This possibility stems from the  fact that the CoRoT-7 system is a rapidly evolving extrasolar
system with very close-in planets.
 
We did our study using numerical integrations for the long term development of
the system, as well as an analytical approach for short time
intervals in the order of years. It turned out the system is quite stable even
in the 3 planet model. A quantitatively new result is the dependence on the difference in
the ascending node on the short time development of \ccc b. The small phase
shift in the dynamical development is not important for the qualitative behaviour for
long time, but essential for the short time evolution of the orbit of \ccc b.

In our determination of the duration of the transit caused by the change of the
inclination of \ccc b we also discussed what kind of orbits of the outer planet(s) would lead to a transit of these planets. 

Despite the constraints given by the incompleteness of the data derived from observations the main conclusion of our study is that after three years of observation of the EPS 
CoRoT-7 we would be able to determine  -- via transit observations from space of the
CoRoT satellite and additional ground based RV -- whether the two (three planets) are on
mutually inclined orbits or whether they have just small inclinations
of the order of  $i<10^{\circ}$. 
%Larger mutual inclinations would result in a change of
%the duration in the order of 1 minutes when the respective inclination of the
%transiting planet would suffer from a slight decrease in the order of 0.3
%caused by the inclination of one or two outer planets. 
More work has to be
done for a more detailed analysis with numerical studies but also with
analytical approaches \citep[e.g.][]{Bor03,Bor07} comparable to those which have been undertaken for the investigation of the change in occultations of eclipsing binaries caused by an additional star.

\begin{acknowledgements}
V. Eybl wants to acknowledge the support from the Austrian FWF project 18930-N16.
\end{acknowledgements}

\bibliographystyle{aa} % style aa.bst
\bibliography{n24a}

\begin{thebibliography}{19}
\expandafter\ifx\csname natexlab\endcsname\relax\def\natexlab#1{#1}\fi

\bibitem[{{Batygin} {et~al.}(2009){Batygin}, {Laughlin}, {Meschiari}, {Rivera},
  {Vogt}, \& {Butler}}]{Bat09}
{Batygin}, K., {Laughlin}, G., {Meschiari}, S., {et~al.} 2009, \apj, 699, 23

\bibitem[{{Borkovits} {et~al.}(2010){Borkovits}, {Csizmadia}, \&
  {Forg\'acs-Dajka}}]{Bor10}
{Borkovits}, T., {Csizmadia}, S., \& {Forg\'acs-Dajka}, E. 2010, {in
  preparation}

\bibitem[{{Borkovits} {et~al.}(2003){Borkovits}, {{\'E}rdi},
  {Forg{\'a}cs-Dajka}, \& {Kov{\'a}cs}}]{Bor03}
{Borkovits}, T., {{\'E}rdi}, B., {Forg{\'a}cs-Dajka}, E., \& {Kov{\'a}cs}, T.
  2003, \aap, 398, 1091

\bibitem[{{Borkovits} {et~al.}(2007){Borkovits}, {Forg{\'a}cs-Dajka}, \&
  {Reg{\'a}ly}}]{Bor07}
{Borkovits}, T., {Forg{\'a}cs-Dajka}, E., \& {Reg{\'a}ly}, Z. 2007, \aap, 473,
  191

\bibitem[{{Gilliland} {et~al.}(2010){Gilliland}, {Brown},
  {Christensen-Dalsgaard}, {Kjeldsen}, {Aerts}, {Appourchaux}, {Basu},
  {Bedding}, {Chaplin}, {Cunha}, {De Cat}, {De Ridder}, {Guzik}, {Handler},
  {Kawaler}, {Kiss}, {Kolenberg}, {Kurtz}, {Metcalfe}, {Monteiro}, {Szab{\'o}},
  {Arentoft}, {Balona}, {Debosscher}, {Elsworth}, {Quirion}, {Stello},
  {Su{\'a}rez}, {Borucki}, {Jenkins}, {Koch}, {Kondo}, {Latham}, {Rowe}, \&
  {Steffen}}]{Gil10}
{Gilliland}, R.~L., {Brown}, T.~M., {Christensen-Dalsgaard}, J., {et~al.} 2010,
  \pasp, 122, 131

\bibitem[{{Hatzes} {et~al.}(2010){Hatzes}, {Dvorak}, {Wuchterl},
  {et~al.}}]{Hat10}
{Hatzes}, A., {Dvorak}, R., {Wuchterl}, G., {et~al.} 2010, {in preparation}

\bibitem[{{Kipping}(2008)}]{Kip08}
{Kipping}, D.~M. 2008, \mnras, 389, 1383

\bibitem[{{L{\'e}ger} {et~al.}(2009){L{\'e}ger}, {Rouan}, {Schneider}, {Barge},
  {Fridlund}, {Samuel}, {Ollivier}, {Guenther}, {Deleuil}, {Deeg}, {Auvergne},
  {Alonso}, {Aigrain}, {Alapini}, {Almenara}, {Baglin}, {Barbieri}, {Bruntt},
  {Bord{\'e}}, {Bouchy}, {Cabrera}, {Catala}, {Carone}, {Carpano}, {Csizmadia},
  {Dvorak}, {Erikson}, {Ferraz-Mello}, {Foing}, {Fressin}, {Gandolfi},
  {Gillon}, {Gondoin}, {Grasset}, {Guillot}, {Hatzes}, {H{\'e}brard}, {Jorda},
  {Lammer}, {Llebaria}, {Loeillet}, {Mayor}, {Mazeh}, {Moutou}, {P{\"a}tzold},
  {Pont}, {Queloz}, {Rauer}, {Renner}, {Samadi}, {Shporer}, {Sotin}, {Tingley},
  {Wuchterl}, {Adda}, {Agogu}, {Appourchaux}, {Ballans}, {Baron}, {Beaufort},
  {Bellenger}, {Berlin}, {Bernardi}, {Blouin}, {Baudin}, {Bodin}, {Boisnard},
  {Boit}, {Bonneau}, {Borzeix}, {Briet}, {Buey}, {Butler}, {Cailleau},
  {Cautain}, {Chabaud}, {Chaintreuil}, {Chiavassa}, {Costes}, {Cuna Parrho},
  {de Oliveira Fialho}, {Decaudin}, {Defise}, {Djalal}, {Epstein}, {Exil},
  {Faur{\'e}}, {Fenouillet}, {Gaboriaud}, {Gallic}, {Gamet}, {Gavalda},
  {Grolleau}, {Gruneisen}, {Gueguen}, {Guis}, {Guivarc'h}, {Guterman},
  {Hallouard}, {Hasiba}, {Heuripeau}, {Huntzinger}, {Hustaix}, {Imad},
  {Imbert}, {Johlander}, {Jouret}, {Journoud}, {Karioty}, {Kerjean},
  {Lafaille}, {Lafond}, {Lam-Trong}, {Landiech}, {Lapeyrere}, {Larqu{\'e}},
  {Laudet}, {Lautier}, {Lecann}, {Lefevre}, {Leruyet}, {Levacher}, {Magnan},
  {Mazy}, {Mertens}, {Mesnager}, {Meunier}, {Michel}, {Monjoin}, {Naudet},
  {Nguyen-Kim}, {Orcesi}, {Ottacher}, {Perez}, {Peter}, {Plasson}, {Plesseria},
  {Pontet}, {Pradines}, {Quentin}, {Reynaud}, {Rolland}, {Rollenhagen},
  {Romagnan}, {Russ}, {Schmidt}, {Schwartz}, {Sebbag}, {Sedes}, {Smit},
  {Steller}, {Sunter}, {Surace}, {Tello}, {Tiph{\`e}ne}, {Toulouse}, {Ulmer},
  {Vandermarcq}, {Vergnault}, {Vuillemin}, \& {Zanatta}}]{Leg09}
{L{\'e}ger}, A., {Rouan}, D., {Schneider}, J., {et~al.} 2009, \aap, 506, 287

\bibitem[{{Libert} \& {Tsiganis}(2009{\natexlab{a}})}]{Lib09b}
{Libert}, A. \& {Tsiganis}, K. 2009{\natexlab{a}}, \aap, 493, 677

\bibitem[{{Libert} \& {Tsiganis}(2009{\natexlab{b}})}]{Lib09a}
{Libert}, A. \& {Tsiganis}, K. 2009{\natexlab{b}}, \mnras, 400, 1373

\bibitem[{{Lovis} {et~al.}(2010)}]{Lov10}
{Lovis}, C. {et~al.} 2010, submitted to \aap

\bibitem[{{Mardling}(2010)}]{Mar10}
{Mardling}, R.~A. 2010, ArXiv e-prints

\bibitem[{{Nagasawa} {et~al.}(2008){Nagasawa}, {Ida}, \& {Bessho}}]{Nag08}
{Nagasawa}, M., {Ida}, S., \& {Bessho}, T. 2008, \apj, 678, 498

\bibitem[{{O'Donovan} {et~al.}(2006){O'Donovan}, {Charbonneau}, {Mandushev},
  {Dunham}, {Latham}, {Torres}, {Sozzetti}, {Brown}, {Trauger}, {Belmonte},
  {Rabus}, {Almenara}, {Alonso}, {Deeg}, {Esquerdo}, {Falco}, {Hillenbrand},
  {Roussanova}, {Stefanik}, \& {Winn}}]{Odo06}
{O'Donovan}, F.~T., {Charbonneau}, D., {Mandushev}, G., {et~al.} 2006, \apjl,
  651, L61

\bibitem[{{Queloz} {et~al.}(2009){Queloz}, {Bouchy}, {Moutou}, {Hatzes},
  {H{\'e}brard}, {Alonso}, {Auvergne}, {Baglin}, {Barbieri}, {Barge}, {Benz},
  {Bord{\'e}}, {Deeg}, {Deleuil}, {Dvorak}, {Erikson}, {Ferraz Mello},
  {Fridlund}, {Gandolfi}, {Gillon}, {Guenther}, {Guillot}, {Jorda}, {Hartmann},
  {Lammer}, {L{\'e}ger}, {Llebaria}, {Lovis}, {Magain}, {Mayor}, {Mazeh},
  {Ollivier}, {P{\"a}tzold}, {Pepe}, {Rauer}, {Rouan}, {Schneider},
  {Segransan}, {Udry}, \& {Wuchterl}}]{Que09}
{Queloz}, D., {Bouchy}, F., {Moutou}, C., {et~al.} 2009, \aap, 506, 303

\bibitem[{{Scuderi} {et~al.}(2010){Scuderi}, {Dittmann}, {Males}, {Green}, \&
  {Close}}]{Scu10}
{Scuderi}, L.~J., {Dittmann}, J.~A., {Males}, J.~R., {Green}, E.~M., \&
  {Close}, L.~M. 2010, \apj, 714, 462

\bibitem[{{Seager} \& {Mall{\'e}n-Ornelas}(2003)}]{Sea03}
{Seager}, S. \& {Mall{\'e}n-Ornelas}, G. 2003, \apj, 585, 1038

\bibitem[{{Stumpf}(1965)}]{Stu65}
{Stumpf}, K. 1965, {Himmelsmechanik Band II. Das Dreik{\"o}rperproblem}
  (Deutscher Verlag der Wissenschaften)

\bibitem[{{Thommes} \& {Lissauer}(2003)}]{ThomLis}
{Thommes}, E.~W. \& {Lissauer}, J.~J. 2003, \apj, 597, 566

\end{thebibliography}
\end{document}